\documentclass[12pt]{article}
\usepackage{graphics,epsfig,rotating}
\begin{document}

\newcommand{\goo}{\,\raisebox{-.5ex}{$\stackrel{>}{\scriptstyle\sim}$}\,}
\newcommand{\loo}{\,\raisebox{-.5ex}{$\stackrel{<}{\scriptstyle\sim}$}\,}

\begin{center}
\large{ 
\bf{ 
On pressure versus density dependence in statistical 
multifragmentation models}\footnote 
{A comment on "Investigating the phase diagram of finite extensive and 
nonextensive systems" by Al.H. Raduta and Ad.R. Raduta, {\it
Phys. Rev. Lett}. {\bf 87}, 202701 (2001)}
}
\end{center}
\vspace{0.1cm}

\begin{center}
\large{
         A.S.~Botvina$^{a,b}$ and I.N.~Mishustin$^{c,d,e}$}

\end{center}
\begin{center}
\small{
 $^a$Gesellschaft f\"ur Schwerionenforschung, D-64291 Darmstadt, Germany\\ 
 $^b$Institute for Nuclear Research, 117312 Moscow, Russia\\
 $^c$Institute for Theoretical Physics, Goethe University, D-60054 
     Frankfurt/Main, Germany\\
 $^d$Niels Bohr Institute, DK-2100 Copenhagen, Denmark\\
 $^e$Kurchatov Institute, Russian Research Center, 123182 Moscow, Russia\\
}
\end{center}

\normalsize
\vspace{0.3cm}

\begin{abstract}
We show that the statistical multifragmentation model with the standard 
parameterization of the free volume predicts a constant 
pressure as density approaches the normal nuclear density. This is contrary 
to the Raduta$\&$Raduta result obtained by disregarding the center of mass
constraint. It is demonstrated that in finite nuclear systems the partitions 
with small number of fragments play an important role. 
\end{abstract}
\vspace{0.15cm}
\hspace{0.8cm} \small{ $PACS$: 25.70.Pq, 24.10.Pa, 24.60-k, 05.70.Fh }

\normalsize
\vspace{0.7cm}

In recent years much efforts have been directed to investigating 
thermodynamical properties of finite nuclear systems (see e. g. 
refs.~\cite{Dag}). 
In the paper \cite{raduta01} the authors study the phase 
diagram of finite systems with a statistical model designed for describing 
nuclear multifragmentation. The authors' model follows the theoretical 
prescriptions developed earlier in refs. [3--5] and
based on the assumption that fragments are 
formed at a low density freeze-out stage. Consequently, the thermodynamical 
characteristics calculated within such kind of models have adequate physical 
meaning at low densities only, e.g. 
$\rho \loo (\frac{1}{2}-\frac{1}{3})\rho_0$ 
($\rho_0 \approx$0.15$fm^{-3}$ is the normal nuclear density), when individual 
fragments are singled out from the surrounding nuclear matter 
\cite{PR95,gross90}. However, the authors of ref. \cite{raduta01} have tried 
to apply their model for the higher densities $\rho \rightarrow \rho_0$ 
by removing the physical assumption that fragments do not overlap. They have 
found that the model predicts a Van der Waals kind of phase 
diagram. In particular, the pressure $P \rightarrow \infty$ at 
$\rho \rightarrow \rho_0$, as seen from Fig.~3 of ref.~\cite{raduta01}. 
In this comment we want to demonstrate that this result can be 
misleading, even under assumption of overlapping fragments. 
 Within the standard statistical treatment of finite systems, 
when the conservation laws are properly implemented, the 
pressure always remains finite, even in the limit $\rho \rightarrow \rho_0$. 

In the statistical models [2--5] the volume 
(density) affects the partition probabilities via so called free volume 
$V_f$ determining the phase space available for the translational motion of 
fragments. It differs from the actual physical volume of the system $V$ 
because of the finite size of the fragments and fragment-fragment interaction.
In the standard canonical description of multifragmentation the free volume 
enters in the partition probabilities through the factor 
$(V_f/\lambda_T^3)^{N-1}$, where $N$ is the number of fragments in the 
partition and $\lambda_T=2\pi \hbar/\sqrt{m_NT}$ is the thermal wavelength 
of nucleons at temperature $T$. 
The exponent $(N-1)$ comes 
from the integration over momenta and coordinates of fragments under
constraints that total momentum and center of mass position are fixed for all
partitions. These constraints are crucial for finite systems in contrast to 
the thermodynamic limit ($N \rightarrow \infty~, V \rightarrow \infty~, 
N/V ={\rm const}$), where this factor is usually taken as 
$(V_f/\lambda_T^3)^N$. 
In the canonical ensemble, for a finite system with the number of nucleons 
$A_0$, the free energy can be represented as 
\begin{equation} \label{fenerg}
F=-T \cdot ln \left ( \sum_{N=1}^{A_0} V_f^{N-1} C_N \right ),
\end{equation}
where $C_N$ is the volume-independent weight of $N$-fragment partitions
which is influenced by combinatorial factors as well as by internal
excitation of fragments and their interaction.  
For simplicity, below we disregard the Coulomb interaction in the system. 
In the microcanonical ensemble the structure of the statistical weights 
with respect to the volume will be practically the same, and thus  our 
conclusions concerning the volume dependence of the pressure in the limit 
$V_f \rightarrow 0$ will not change. 
Though the weights $C_N$ can significantly vary from 
partition to partition, they are {\it finite} for any finite system. 
Therefore, the behavior of the free energy in the limit $V_f \rightarrow 0$ 
is similar for different assumptions on the fragments' relative motion 
in the freeze-out volume. 
To illustrate our point we use the excluded volume approximation $V_f=V-V_0$,
where $V_0=A_0/\rho_0$ is the normal nuclear volume of all fragments.
In this approximation one gets straightforwardly:
\begin{equation} \label{press}
P=-\frac{\partial F}{\partial V}=
T \frac{\sum_{N=2}^{A_0} (N-1) V_f^{N-2} C_N}
{\sum_{N=1}^{A_0} V_f^{N-1} C_N}.
\end{equation} 
Now it is obvious that in the limit $V \rightarrow V_0$ the pressure goes to
a constant, 
$P \rightarrow T \cdot (C_2/C_1)=const$. 
It is interesting to note that a similar behavior for this kind of model 
was also found in the thermodynamic limit \cite{bugaev}. In this case 
constant parts of pressure isotherms appear in the coexistence region of 
liquid and gaseous phases. And the liquid phase is represented by an infinite 
cluster which in a finite system would correspond to the compound-like 
nucleus. 

We believe that the main 
reason of the $P \rightarrow \infty$ behavior obtained in ref. 
\cite{raduta01} is in disregarding the center of mass constraint 
\footnote{Because of this and other approximations used in the model 
\cite{raduta01}, depicting it as "exact" in comparison with other 
models does not seem justified.}. The unconstrained integration over 
coordinates of all fragments 
in the freeze-out volume results in a factor 
proportional to $V_f^N$. 
If in eqs.~(\ref{fenerg}) and (\ref{press}) 
we take $V_f^N$ instead of $V_f^{N-1}$, 
we get formally $P \sim 1/V_f$ at $V \rightarrow V_0$, 
i.e. $P \rightarrow \infty$. 
One can easily see that $V_{free}$ adopted 
in ref.~\cite{raduta01} (their eq. (4)) 
gives the same limiting behavior as $V_f^N$. 

\begin{figure} [tbh]
\vspace{-1cm}
\hspace{1cm}
\includegraphics[width=14cm]{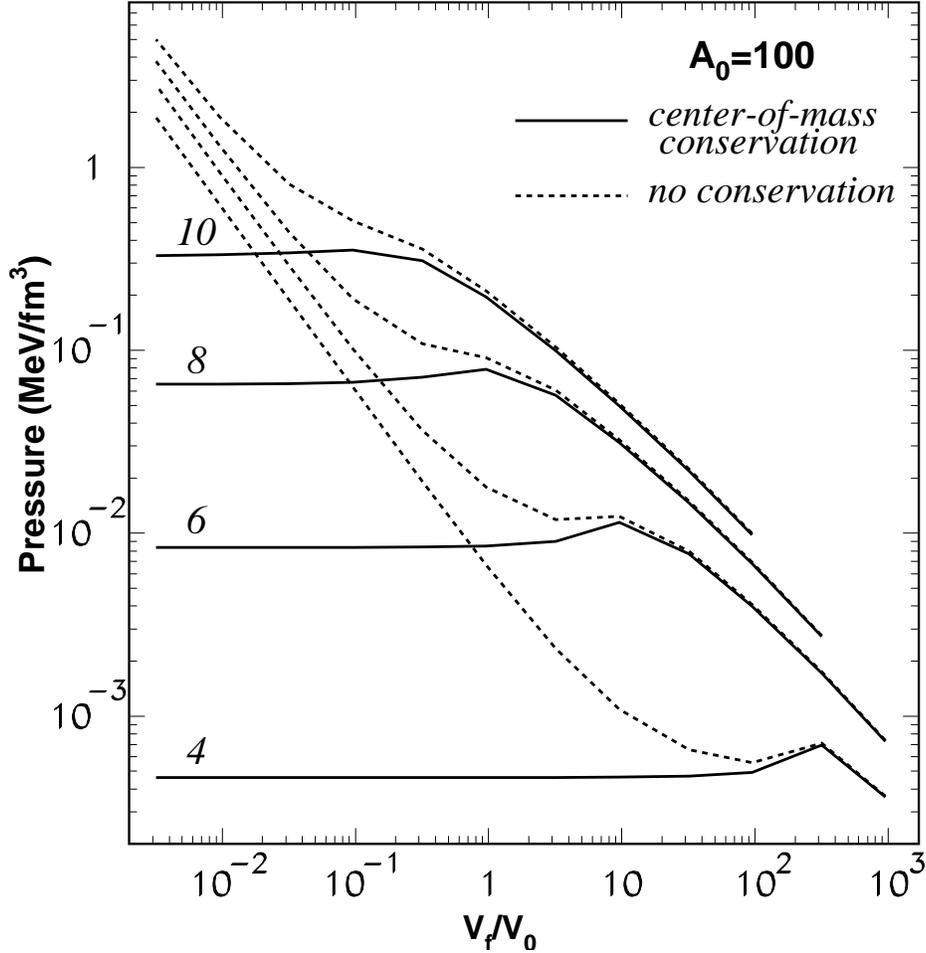}
\caption{\small{
Pressure versus free volume (in units of normal 
nuclear volume) in multifragmentation of one component system with 
$A_0$=100 nucleons calculated within the canonical ensemble. 
Solid lines are calculations taking into account the center-of-mass 
constraint in the fragmentation statistical weights, dashed 
lines correspond to a similar calculation but without this constraint. 
The temperatures (in MeV) are given by numbers at the curves.
}}
\end{figure}

For illustration, in Fig.~1 we show the phase diagram obtained with 
the canonical calculations for 
partitioning a one-component system containing $A_0$=100 nucleons into 
all possible fragments characterized by the mass number $A$ only. 
Contributions of all 190569292 partitions were calculated directly, in oder to 
achieve the best accuracy \cite{botmis}. 
We adopt the liquid-drop description for individual fragments disregarding 
the Coulomb interaction and internal excitation of fragments. 
The statistical weight of a partition 
with $N$ fragments with individual multiplicities $N_A$ is taken as 
\begin{equation} \label{weight}
W(\{N_A\}) \sim \left( \frac{V_f}{\lambda_T^3}\right)^{N-1} \prod_A
\frac{1}{N_A!}\left[A^{3/2}\exp{\left(\frac{B(A)}{T}\right)}\right]^{N_A},
\end{equation}
where $B(A)=a_V\cdot A-a_S\cdot A^{2/3}$ is the liquid-drop binding energy 
of fragment $A$ with parameters $a_V\approx a_S\approx 16$ MeV. 
In this case the pressure is $P=T\cdot (\langle N \rangle -1)/V_f$, 
where $\langle N \rangle$ is the mean fragment multiplicity. 
It is seen that at large $V_f$ the phase diagram is consistent with 
expectations for a gas system. Here the system disintegrates 
into many fragments and the fragment mass distribution falls off nearly 
exponentially with $A$. By decreasing $V_f$ we move into region where the 
partitions with low fragment multiplicity dominate. The mass 
distribution turns into "U-shape"-like one, which is associated with the phase 
transition. Namely in this region the pressure has a slight "backbending" 
(at low temperatures) 
and then approaches a constant. Formally this behavior at small volumes sets 
in because probabilities of the channels with $N>1$ are suppressed by factors 
$\propto V_f^{N-1}$. The compound nucleus (a partition with $N$=1) 
dominates in this case, however, it {\it does not} 
contribute to the pressure. This is a trivial consequence of the 
conservation laws: the compound nucleus is at rest in its center of mass 
frame. For $N>1$ 
only relative momenta and positions of fragments have physical meaning. 

We have also simulated the effect of non-conservation of the center-of-mass
by introducing an additional factor $\propto V_f$ in the weights of 
all partitions in the formula~(\ref{weight}). This effect is shown by dashed 
lines in Fig.~1, corresponding to the pressure 
$P=T\cdot \langle N \rangle/V_f$. 
This means that compound nucleus can "move" and ``exert a pressure'' 
in the freeze-out volume, and as a result $P \rightarrow \infty$ 
at $V_f \rightarrow 0$. Moreover, in this case the 
phase diagram looks like a Van der Waals one, that in fact does not 
correspond to the physical content of the model.  Even in the region of 
large volumes relevant for the model this effect can considerably change 
many thermodynamical characteristics, such as the critical temperature or 
the caloric curve at constant pressure. 

This example shows that phase diagrams of finite systems depend 
sensitively on physical assumptions adopted in the model. 
In particular, in the study of a liquid-gas type phase transition a careful 
treatment of the partitions 
with small multiplicities becomes extremely important. 
However, in order 
to investigate realistically the phase diagram at $\rho > \frac{1}{2} \rho_0$ 
it is necessary to introduce a really {\it new physics} in the 
statistical models, 
e.g. instead of the picture of 
individual fragments surrounded by the nucleon gas to consider the 
bubbles of the nuclear gas inside the nuclear matter.

\end{document}